\documentclass[iop]{emulateapj}
\newcommand{\myemail}{maggie@physics.mcgill.ca}

\newcommand{\psr}{PSR~B1509$-$58}
\newcommand{\kes}{PSR~J1846$-$0258}

\newcommand{\tempo}{{\tt{TEMPO}}}
\newcommand{\nudotdotdot}{{\ifmmode\stackrel{\bf \,...}{\textstyle \nu}\else$\stackrel{\,\...}{textstyle \nu}$\fi}}
\newcommand{\rxte}{\textit{RXTE}}
\newcommand{\xte}{\textit{RXTE}}
\newcommand{\degrees}{^{\circ}}

\usepackage{graphicx}
\usepackage{natbib}

\shorttitle{Long-term X-ray Monitoring of \psr}
\shortauthors{Livingstone et al.}

\begin{document}
\title{Long-term X-ray Monitoring of the Young Pulsar \psr}

\author{Margaret A.~Livingstone \altaffilmark{1},
Victoria M.~Kaspi}
\affil{Department of Physics, Rutherford Physics Building, 
McGill University, 3600 University Street, Montreal, Quebec,
H3A 2T8, Canada}

\altaffiltext{1}{\myemail}

\clearpage

\begin{abstract}
It has long been thought that the pulsed X-ray properties of
rotation-powered pulsars are stable on long time scales. However, 
long-term, systematic studies of individual sources have been 
lacking. Furthermore, dramatic X-ray variability has now been observed
from two pulsars having inferred 
sub-critical dipole magnetic fields. 
Here we present an analysis of the long-term pulsed X-ray properties of the
young, energetic pulsar \psr\ using data from the
{\textit{Rossi X-ray Timing Explorer}}. 
We measured the 2--50\,keV pulsed flux for 14.7\,yr of X-ray observations and
found that it is consistent with being constant on all relevant time
scales,
and place a 3$\sigma$ upper limit on day-to-week variability of  $<$28\%. 
In addition, we searched for magnetar-like X-ray bursts in all
observations and found none, which we use to constrain the measurable burst rate
to less than one per 750\,ks of observations. 
We also searched for variability in the
pulse profile and found that it is consistent with being stable
on time scales of days to decades.
This supports the hypothesis 
that X-ray properties of rotation-powered
X-ray pulsars can be stable on decade-long time scales.
In addition, we extend the existing timing solution by 7.1\,yr to a total of
28.4\,yr and report updated values of the braking index, 
$n=2.832\pm0.003$ and the second braking index, $m=17.6\pm1.9$. 
\end{abstract}

\section{Introduction}
\label{sec:intro}
The X-ray emission properties of non-accreting neutron stars are today
known to be extremely diverse.
There are multiple classes of neutron stars identified --
rotation-powered pulsars (RPPs), Anomalous X-ray
Pulsars (AXPs), Soft Gamma Repeaters (SGRs), thermally emitting
isolated neutron stars, central compact objects --
to name a few (see Kaspi 2010, for a review), \nocite{kas10} 
with the basis for
classification of a given neutron star into
these different classes often being its X-ray emission properties.

Anomalous X-ray Pulsars (AXPs) and Soft Gamma Repeaters (SGRs)  are
now well established as
being extremely variable in the X-ray range, showing a wide variety of
often dramatic bursting behavior during
active periods, in addition to lower-level flux and profile
variability during periods of relative quiescence
(see Woods \& Thompson 2006, Kaspi 2007, Mereghetti 2008, and Rea \&
Esposito 2011, for reviews). \nocite{wt06,kas07,mer08,re11} 
This variability is ascribed to activity both within the stellar
interior as well as in the magnetosphere, both owing ultimately to the
decay of these objects' extremely high magnetic fields ($B > B_{QED}
\equiv 4.4 \times 10^{13}$~G), particularly
in the context of the now widely accepted magnetar model
\citep{td96a,tlk02}.

Interestingly, two recent results  have lowered the
estimated dipolar field strength necessary for magnetar-like
activity to well below
$B\simeq 10^{14}$\,G \citep[e.g.][]{tlk02}. The young RPP \kes\ with
$B=4.9\times10^{13}$\,G experienced a magnetar-like outburst in 2006
\citep{ggg+08,ks08a}, while the SGR~0418+5729, discovered via an
outburst in 2009, has an upper limit on its dipole field of
$B<7.5\times10^{12}$\,G \citep{ret+10}. 
Thus, the smallest possible $B$ field
needed to power a magnetar-like outburst is not currently known.
These recent discoveries highlight the fact that the
link between magnetars and RPPs is poorly
understood. In particular, it suggests that the X-ray emission of some
RPPs may be more variable than previously thought.
Indeed, a study by \citet{pp11} argues that bursting behavior
may be generically possible in all
RPPs, only more common in those that are young and
with high $B$.

Systematic examinations of the X-ray variability properties of
RPPs in order to look
for other examples of bursting behavior or flux changes has been
hampered by the difficulty in obtaining  a well sampled data set 
with a single, well calibrated instrument over a long
time span.  Rotation-powered pulsars are generally too faint for
detection with all-sky monitors.

PSR B1509$-$58 is a young, energetic pulsar that was discovered with
the {\it Einstein} satellite 
\citep{sh82}.  It has period 150\,ms, a high spin-down luminosity of
$\dot{E} = 1.7 \times 10^{37}$\,erg\,s$^{-1}$,
and a relatively large inferred\footnote{Note
that the dipolar magnetic field for pulsars is 
$B=3.2\times10^{19}\sqrt{P \dot{P}}$\,G.} dipolar magnetic field of 
$B=1.5 \times 10^{13}$\,G, larger than that for the magnetar with the smallest estimate of
$B<7.5\times10^{12}$\,G \citep{ret+10}.
\psr's high spin-down rate and lack of glitches have allowed the
measurement of higher-order 
spin parameters, in particular, the second and third frequency
derivatives ($\ddot{\nu}$ and $\nudotdotdot$).
These provide a determination of the braking index, 
$n=\nu\ddot{\nu}/{\dot\nu}^2 = 2.839\pm0.003$, and second braking
index, $m= \nu^2\nudotdotdot/{\dot\nu}^3= 18.3\pm2.9$ \citep{kms+94,lkgm05},
allowing tests of the standard pulsar spin-down model \citep[e.g.][]{bla94}.

PSR B1509$-$58 presents an excellent opportunity for a systematic,
long-term study of a RPP's X-ray emission properties, since it has been observed
regularly since 1996 with 
the Proportional Counter Array (PCA) aboard the {\it Rossi X-ray
Timing Explorer (RXTE)} as a timing
calibration source. Here we present a search  of 14.7 yr of {\it RXTE}  data
of \psr\ for pulsed flux variations, pulse
profile variations, and magnetar-like bursts. We
also extend the timing solution for the pulsar
by 7.1 yr, increasing the total observing time to
28.4 yr. We use these new data to refine the measurements
of $n$  and $m$, and further quantify the variability of $n$.

\section{Observations}
\label{sec:obs}

\psr\ data were obtained with the Proportional
Counter Array \citep[PCA;][]{jsg+96,jmr+06} on board \textit{RXTE}. 
The PCA consists of five collimated
xenon/methane multi-anode proportional counter units (PCUs) operating
in the 2\,--\,60~keV range, with a total effective area of approximately
$\rm{6500\,cm^2}$ and a field of view of $\rm{\sim 1^o}$\,FWHM.
We downloaded 290 archival \rxte\ observations of \psr\ 
from
the HEASARC archive\footnote{http://heasarc.gsfc.nasa.gov/docs/archive.html}. 
All data were collected in
the GoodXenon mode, which records the arrival time
(1 $\mu$s resolution) and energy (256 channel resolution) of each
unrejected event. 
The observations span 7.1\,yr from 
MJD~50148--55221 (1996 March 6 -- 2010 November 21). 

The data reduction and preparation is significantly different 
for the timing and pulse profile analyses than for the 
pulsed flux analysis. Details for each are included in the following
sections. 

\section{Timing analysis}
\label{sec:timing}
Our timing analysis of \psr\ follows the common phase-coherent approach, 
in which we account for each rotation of the pulsar. 
Pulse times of arrival are fitted with a Taylor expansion of pulse 
phase \citep[see, for example, ][]{mt77}, allowing the extraction of
precise spin parameters (spin-frequency $\nu$, and frequency
derivatives $\dot\nu$, $\ddot\nu$, and $\nudotdotdot$).

In order to produce pulse times-of-arrival (TOAs), 
we extracted photons from all three xenon layers of each PCU in the
2~--~30\,keV energy range (or 4~--~72 channel range).
Data from the individual PCUs were merged and binned at 1/1024\,s resolution.
The time series were reduced to barycentric dynamical time (TDB)
at the solar system barycenter using the known position from radio
interferometry \citep[J2000 RA = $15^{\rm{h}}$~$13^{\rm{m}}$\,$55\fs62$, Dec
$=-59\degrees$\,$08\arcmin$\,$9.00\arcsec$;][]{gbm99} and the JPL DE200 solar
system ephemeris.

When more than one observation occurred in a 24-hr period, we combined
the data from multiple observations into a single time series in order
to produce better TOAs and having smaller uncertainties.
Each time series was folded with 64 phase bins using an initial ephemeris
from \citet{lkgm05}. Resulting pulse profiles were cross-correlated in the
Fourier domain with a high signal-to-noise-ratio template. We implemented a
Fourier domain filter by using six harmonics in the cross-correlation
procedure. Cross-correlation produces an average TOA
for each observation with a typical uncertainty of $\sim$0.5\,ms. 
TOAs were fitted phase coherently with the pulse timing software package
{\tt{TEMPO}}\footnote{www.atnf.csiro.au/people/pulsar/tempo/}; timing 
residuals and refined spin parameters are produced as output.
Spin parameters for all 14.7\,yr of X-ray data are given in
Table~\ref{table:coherent}.

\begin{figure}
\plotone{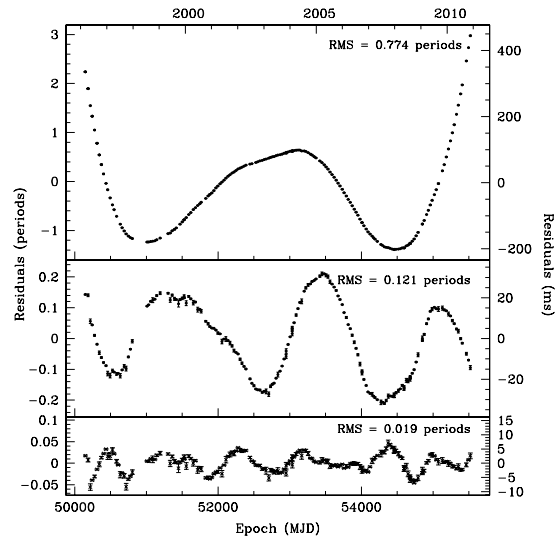}
\figcaption[1509_xray_resids.eps]{\rxte\ timing residuals for \psr\
spanning 1996 March -- 2010 November.
The top panel has $\nu$, $\dot\nu$, and $\ddot\nu$ fitted;
the middle panel has a quartic polynomial (i.e. $\nudotdotdot$
fitted), and
the bottom panel has 12 derivatives fitted. Spin parameters from a
fully phase-coherent fit are given in Table~\ref{table:coherent}.
\label{fig:resids}}
 \end{figure}

The top panel of Figure~\ref{fig:resids} shows timing residuals with
pulse frequency, $\nu$, frequency derivative, $\dot\nu$, and
second frequency derivative, $\ddot\nu$, fitted. No sudden jumps were
found during the phase connection process, nor were any cusps
identified in the timing residuals, indicating that no glitches
occurred during the time spanned by the \rxte\ observations. 
The middle panel shows residuals with 
$\nudotdotdot$ also fitted. \psr\ is one of two pulsars with
deterministic measurements of $\ddot\nu$ and $\nudotdotdot$
\citep{kms+94,lkgm05}, however, after the removal of 
all deterministic parameters, significant structure remains in the
timing residuals, as is visible in the middle panel. This can be attributed
to timing noise, a low-frequency or ``red'' noise process that is 
a common but poorly understood phenomenon in pulsars
\citep[and previously discussed in detail for this pulsar
in][]{lkgm05}. {\tt{TEMPO}} has the
capability to fit up to 12 frequency derivatives; the
results of this fit are shown in the bottom panel of
Figure~\ref{fig:resids}.
The presence of timing noise is known to contaminate measurements of
higher order spin parameters such as $n$ and $m$ and complicates the
proper estimation of uncertainties. One method of
measuring determinsitic parameters and estimating their true
uncertainites in the presence of timing noise is discussed in the following section. 

\normalsize

\subsection{Partially phase-coherent timing analysis}
\label{sec:partial}

One method of finding reliable values for $n$ and $m$ in the presence 
of timing noise is to use a piece-wise approach, deemed ``partially coherent'' timing
analysis. The data are broken into subsets, each of which is
fitted phase-coherently with $\nu$, $\dot\nu$, and $\ddot\nu$. For 
each subset, we require that no 
structure be visible in the resulting timing residuals. 
The amount of data included in
each subset is determined only by this requirement, so is allowed to
vary. On average, the time span in a subset is $\sim$500\,days. 
The
date ranges spanned by these subsets and the resulting ephemerides are
given in Table~\ref{table:partial} (and are used for our pulsed flux
analysis, see Section~\ref{sec:flux}). In order to better sample the
available data, we created additional subsets that overlapped
the original subsets by $\sim 1/2$. Overlapping subsets are useful since 
short coherent timing solutions produce fitted parameters dominated by the
end points \citep[e.g.][]{lkgm05}. Figure~\ref{fig:partial} 
shows all measurements of $\ddot\nu$ both from the current analysis, as well as
previous measurements determined using the same method but 
from both radio and X-ray timing data reported in \citet{lkgm05}. 

We performed a weighted least squares
fit to measure $\nudotdotdot$ and a bootstrap error analysis to 
better estimate the uncertainties, since the data are known to be
contaminated with red noise leading to formal uncertainties that
underestimate the true uncertainties \citep{efr79}. 
This analysis results in $\nudotdotdot = -1.21(13) \times
10^{-31}$\,s$^{-4}$, in agreement with the previously reported value of 
$\nudotdotdot = -1.28(21)\times10^{-31}$\,s$^{-4}$, based on 21.3\,yr
rather than 28.4\,yr of data, but calculated in the same way. 

A value of $\nudotdotdot$ can be used to find a measurement of
the ``second'' braking index, $m$, via $m=\nu^2 \nudotdotdot/{\dot\nu}^3$.
This parameter, a higher order analog to the braking index, $n$, provides a
unique test of pulsar spin-down models \citep[e.g.][]{br88}. 
The above value of $\nudotdotdot$ corresponds to a value of 
$m = 17.6\pm1.9$, smaller than the value from 21.3\,yr of data
reported in \citet{lkgm05} of $m=18.3\pm2.9$, but not significantly so. 

\begin{figure}
\plotone{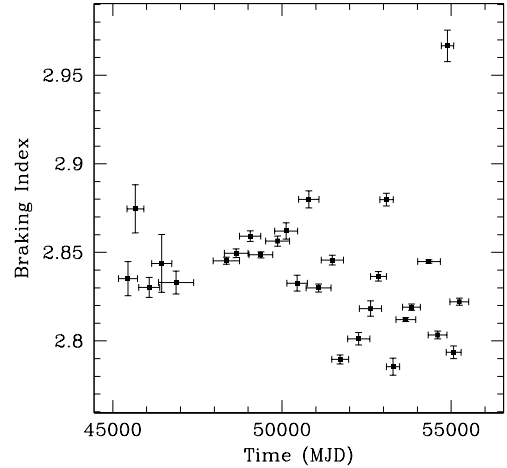}
\figcaption[]{Braking indices from short phase-coherent timing
solutions. Data prior to MJD~53000 were previously reported in
\citep{lkgm05}. The weighted average is $n=2.832\pm0.003$.
At the 2$\sigma$ level, there is no linear change over 28.4\,yr.
Significant scatter is
apparent in the data; the root-mean square
variation from the mean $n$ is at
the 3.5\%-level. \label{fig:braking}}
\end{figure}

We used the same timing solutions to calculate $n$ for each data
subset as shown in Figure~\ref{fig:braking}. 
The weighted average value, with the uncertainty calculated from a
bootstrap analysis is $n=2.832\pm0.003$, smaller than the previously
reported value of $2.839\pm0.003$ at the 1.6$\sigma$ level.
In order to look for a linear
change in $n$ over the entire 28.4\,yr of observations, we 
performed a weighted linear least
squares fit to the data. This results in a slope of
$\Delta{n}/\Delta{t} = (-4\pm2)\times10^{-6}$, which 
is consistent with no secular change in $n$ at the $2\sigma$ level.
While the long-term $n$ is relatively stable, short-term variability
is visible as scatter in the individual measurements, which can
most likely be attributed to timing noise, since no discrete jumps in
$\nu$, i.e. glitches, were found. This amounts to a maximum variation
from the average value of $4.8\pm0.3\%$, and a root-mean-square variation from 
the mean of 3.5\%. 
Because of the large scatter in the measurement of $\ddot\nu$ near
MJD~55000, which is significantly larger than the mean, we further
verified that no glitch occurred in this time period. 

\normalsize

\section{Pulsed flux analysis and results}
\label{sec:flux}

In order to determine the pulsed flux of \psr\ at each epoch, 
for each observation, we created a phase resolved spectrum 
with 8 phase bins, for each PCU. 
Each observation was folded with the software package
{\tt{fasebin}}\footnote{http://heasarc.nasa.gov/docs/xte/recipes/fasebin.html} 
using the appropriate short coherent ephemeris described in
Section~\ref{sec:partial} and listed in Table~\ref{table:partial}.
We used these short ephemerides instead of the global solution in part 
because {\tt{fasebin}} can take a maximum of two frequency derivatives
for an
input ephemeris. For \psr, using the global two-derivative ephemeris
produces very poor pulse profiles because of the effect of the
deterministic $\nudotdotdot$ and timing noise, whereas the profiles
produced from short ephemerides are consistently good. 
A phase-averaged spectrum was also created in order to build a response
matrix for each PCU. For each observation, we combined phase-resolved
spectra and response matrices for all PCUs that were on for $>95\%$ 
of the observation.

From each combined phase-resolved spectrum, we created a pulse profile with 8 phase 
bins in the 2--50\,keV energy range. The minimum bin of each profile
was used to set the background level, and was subtracted from all 
bins of the phase-resolved profile. 

After background subtraction, we combined the spectra for all phase
bins into a single spectrum 
and fitted the result using
{\tt{XSPEC}}\footnote{http://xspec.gsfc.nasa.gov/ Version 11.3.1}.
For each observation, we used a simple power-law model with a fixed spectral
index of $\Gamma=1.19$ and fixed $N_H=8.6\times10^{21}$\,cm$^{-2}$
\citep{cmm+01}, fitting only for the flux. 
The resulting fit $\chi^2$ values showed that this model fit all of 
our data well. 

For short exposure times, there is an observed anti-correlation between the measured pulsed
flux and the source exposure time. This is a statistical effect
present for all sources and disappears when sufficient counts are
detected, i.e. for longer exposure times. 
While one might expect that the uncertainties on pulsed flux
measurements would simply increase for short exposures, in addition,
the measured pulsed flux is biased upwards. 
This arises from Poissonian variations, 
which lead to the minimum bin
being biased downward when there are few total counts, thus to 
an under-estimate of the background and an over-estimate of the
pulsed flux. Because the number of active
PCUs changes from observation to observation, 
for convenience, we define an ``effective'' observation time of 
$T_{\rm{eff}} = N_{\rm{pcus}}  T_{\rm{obs}}$, where 
$N_{\rm{pcus}}$ is the number of PCUs on during an observation    
and $T_{\rm{obs}}$ is the on-source time. We compared the measured flux
values to $T_{\rm{eff}}$ and found a small bias to be present. This bias was
eliminated by excluding all observations having
$T_{\rm{eff}} < 6$\,ks, which amounts to 34\% of the data.

\begin{figure}
\plotone{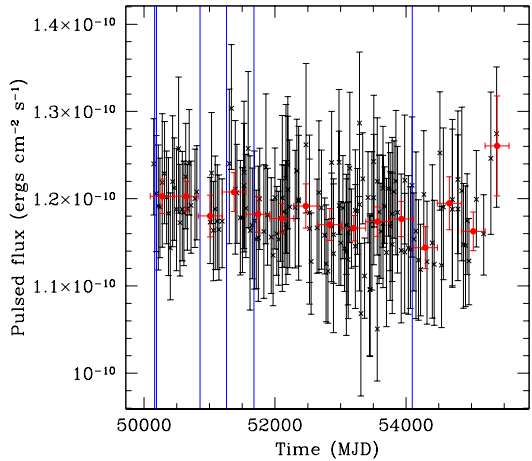}
\figcaption[1509flux.eps]{Absorbed
pulsed flux measurements for \psr\
determined for observations with $T_{\rm{eff}}>6$\,ks. Vertical lines
mark
dates of epoch changes in \rxte. Filled circles (colored red in the
online version) show the average pulsed
flux over year-long periods. The average pulsed flux
in the 2--50\,keV energy range is
$1.181(5)\times10^{-13}$\,erg\,cm$^{-2}$\,s$^{-1}$.
A weighted least-squares fit to the data gives a slope of
$(-7.4\pm3.8)\times 10^{-16}$\,erg\,cm$^{-2}$\,s$^{-1}$\,day$^{-1}$,
indicating that the pulsed flux is consistent with having no linear
trend at the 1.9$\sigma$ level. \label{fig:flux} }
\end{figure}

Figure~\ref{fig:flux} shows all measurements of the pulsed component of
the flux of \psr\ for observations with $T_{\rm{eff}}>6$\,ks. Vertical
lines delimit each of the \rxte\ epochs (1, 2, 3, 3a, 4) and the losses of
the propane layers in PCU0 (also referred to as epoch 5) and
PCU1\footnote{http://heasarc.gsfc.nasa.gov/docs/xte/pca\_history.html
http://heasarc.gsfc.nasa.gov/docs/xte/e-c\_table.html}.
We verified that
there are no significant correlations between the pulsed flux and any of
the gain changes or propane layer loss events. We also measured the
pulsed count rate for each observation and note that 
correlations exist between the pulsed count rate and gain
changes and propane layer losses. In addition, the pulsed count
rate decreases gradually by $\sim$10\% over the lifetime of \rxte, 
associated with the long-term degredation of
the PCA. For these reasons, we report only the pulsed 
flux values, despite somewhat larger uncertainties. 
We note that  \psr\ is listed as a calibration
source for \rxte\ but has only been used as a timing 
calibrator and not a flux calibration 
source\footnote{See, for example, http://heasarc.nasa.gov/docs/xte/abc/time.html}.    

\normalsize

The weighted average 2--50\,keV absorbed pulsed flux for all 14.7\,yr is
$1.181(5)\times10^{-10}$\,erg\,cm$^{-2}$\,s$^{-1}$, with a $\chi^2$
value of 52.4 for 150 degrees of freedom. A weighted linear 
least-squares fit to the pulsed flux values $f$ gives
$\Delta{f}/\Delta{t} =
(-7.4\pm3.8)\times10^{-16}$\,erg\,cm$^{-2}$\,s$^{-1}$\,day$^{-1}$,
that is, consistent with having no linear trend at the 1.9$\sigma$ level.
The maximum variation of the pulsed flux of a single observation
from the average corresponds to a variability of
(10$\pm$6)\%. At the
3$\sigma$ level, we can rule out changes on day-to-week time scales
of greater than 28\% from the average pulsed flux value.
We also studied the flux stability on monthly and yearly time scales by
creating weighted averages of individual flux measurements. Average
flux values for each \rxte\ cycle are shown on Figure~\ref{fig:flux} as
filled circles (colored red in the online version) 
On monthly time scales, at the 3$\sigma$ level, we rule out variation
from the average larger than 20\%, while on yearly time scales, the 
3$\sigma$ upper limit on variability is 21\%.

\section{Pulse profile analysis}
\label{sec:profiles}
In order to search for variability in the X-ray pulse profile of \psr,
we folded each 2--30\,keV time series (as described for the timing
analysis in Section~\ref{sec:timing})
with the appropriate ephemeris given
in Table~\ref{table:partial}. 
Previous studies have examined the pulse profile as a function of 
energy in detail and found
that the pulse shape does not change appreciably in the \rxte\ energy range
\citep{rjm+98,mbg+97}, hence we do not repeat this analysis here and concentrate instead on
the possibility of the profile changing with time. 

\begin{figure}
\plotone{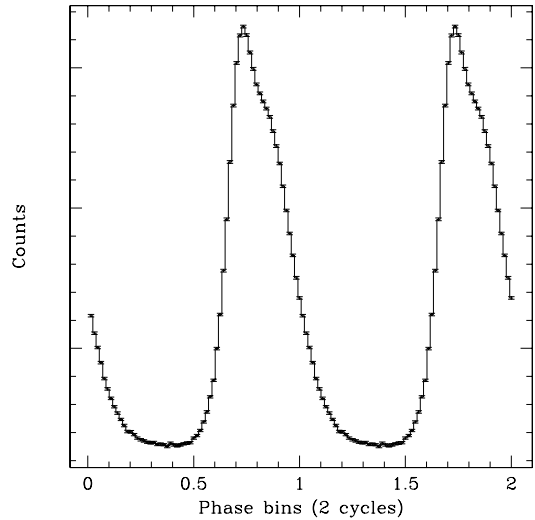}
\figcaption[test_grand_prof.eps]{Summed 64-bin pulse profile
for \psr\ for all 14 Cycles of \rxte\ data for the 2--50\,keV
energy range. We found no significant difference
between individual pulse profiles and the average template, as
discussed in Section~\ref{sec:profiles}.
\label{fig:profile}}
\end{figure}

First, we analyzed the profiles for individual observations in
order to search for variability on time scales of days to months (i.e.
the range of time spans between individual observations). 
We created a template profile by summing all aligned pulse profiles, shown in 
Figure~\ref{fig:profile}. We subtracted a DC offset from both the
individual profiles and template, 
then normalized the template to each
individual profile, subtracted them and 
found the residuals between the two profiles. 
We calculated the reduced
$\chi^2$ of the resulting residuals and in each case, we see no evidence for
pulse profile variability.  To quantify the lack of profile changes, we calculated the
root-mean-square of the profile residuals as a percentage
of the number of counts in each DC-offset subtracted profile and found
a 3$\sigma$ upper limit on variations of 2.1$\%$. 

\normalsize

Next, we created a single profile for each \rxte\ observing 
cycle (spanning $\sim$1-yr each). 
In order to check whether there were any small-scale changes in the
pulse profile, we compared each profile against the
previously described template profile. Repeating the analysis
described above, again we found no evidence for variability. 
For the yearly profiles, we found a 3$\sigma$ upper limit on profile
changes of less than 0.1\%. 

We also compared the profile from each of the 14 \rxte\ cycles with 
that of the previous and subsequent cycle profiles.
We normalized each profile and then created residuals between pairs of
profiles. We then performed a $\chi^2$ test on each set of residuals.
In every case, ${\chi^2}_\nu<1$, i.e. we detected no evidence of long term
changes in the pulse profile. In order to have maximum sensitivity to
changes on the longest possible time scales, we also repeated the
analysis by comparing the Cycle 1 profile with the Cycle 14 profile.
Again, we found that ${\chi^2}_\nu<1$.

\section{Burst search}
\label{sec:bursts}

To search for magnetar-like X-ray bursts from \psr, for each observation
we created a time series for each active PCU from GoodXenon
data, selecting events in the
2~--~20\,keV range and from all three detection layers 
\citep[the same energy range as selected for similar searches for
X-ray bursts from AXPs, e.g.][]{gkw04}. 
This was repeated several times with different time resolutions,
specifically for 15.625\,ms, 31.25\,ms, 62.5\,ms and 125\,ms, to provide 
sensitivity to bursts on a hierarchy
of time scales. The procedure used here was based on that described
in \citet{gkw04}. 

By this algorithm, bursts are identified by comparing the count rate
in the i$^{th}$ bin to the average count rate. Because the background
rate of the PCA typically varies over a single observation, we
calculated a local mean around the i$^{th}$ bin. We then
compared the count rate in the i$^{th}$ bin to the local average.
If the count rate in a single bin was larger than the local
average, the probability of such a count rate occuring by chance was
calculated. Then the probability of the count rate in the
corresponding bin in the other active PCUs was also calculated
(whether or not the count rate in that bin was greater than the local
average). If a PCU was off during the bin of interest, its probability
was set to 1. We then found the total probability that a burst was
observed. If the total probability of an event was $P_{i,{\rm{tot}}} \le 0.01/N$ 
(where $N$ is the total number of time bins searched), it was flagged
as a burst. 

We searched a total of 751\,ks of \rxte\ data using the above
described process and found no X-ray bursts from the direction of \psr. 

Unlike the AXPs, \psr\ has a bright PWN which contributes a large
unpulsed background to the 1$\degrees$
field-of-view of the PCA. Because of this larger background, the
sensitivity to AXP-like X-ray bursts could be significantly smaller
than for the AXPs. Thus, in order to determine the fluence of bursts
that would have been detectable in these data, we added simulated bursts to real
\rxte\ time series of \psr. Each simulated burst was confined to a
single time bin and spread between the operational 
PCUs. We added many bursts to each time series, each separated by
sufficient time that only one burst would ever be included in a
calculation of the running mean. Many bursts were added to each time
series in order to include the effect of the
variation of the background count rate, largely owing to the variable particle
background during the \rxte\ orbit. We repeated this
analysis for each time resolution searched. 
Unsurprisingly, we found that the minimum burst fluence required to 
detect 100\% of the simulated bursts is a function
of both the number of operational PCUs and width of the time bins. 

In terms of detected burst count rates, the smallest detectable bursts are
similar to the smallest bursts detected from the AXP~1E~2259+586 in
its large 2002 outburst \citep{gkw04}. 
For example, for 31.25\,ms time resolution, 100\% of simulated bursts
are detected if the burst consisted of 8 or more counts per PCU or
more than 25 counts/burst total (i.e. at least 25 counts per 31.35\,ms
time bin).
In order to quantify this in terms of burst
fluences, we assumed a burst spectrum of a
simple power law with spectral index of $\Gamma=1.35$, as measured for
the 80 bursts found from the AXP 1E~2259+586 \citep{gkw04}. 
For 62.5\,ms time resolutions, we found that burst fluences of
$\ge 1.5 \times 10^{-11}$\,erg\,cm$^{-2}$ (2--20\,keV)
would have been detected, an order of magnitude less than the 
smallest burst fluence detected from \kes\ in its 2006 outburst
\citep{ggg+08}.  
Detectable burst rates and fluences for all searched time resolutions
are given in Table~\ref{table:bursts}, and are valid in the absence of
instrumental flares. Thus, we can confidently exclude
the possibility that AXP-like X-ray bursts with fluences larger than
the limits given in Table~\ref{table:bursts} were present in the \rxte\ \psr\
data. 

\section{Discussion}
\label{sec:discussion}

We searched for several types of X-ray variability from the
young, high-$B$ field pulsar \psr\ and found that the pulsar is stable
in each way studied. This establishes a baseline for the quiescent
pulsed flux of \psr\ in case of a future magnetar-like outburst. 
We establish a 3$\sigma$ upper limit on pulsed flux
variations of less than 28\% from the average,
ruling out pulsed flux changes similar to those seen in the magnetars and
\kes. For example, during the 2006 outburst of \kes, the 2--60\,keV pulsed
flux as measured with \rxte\
increased by a factor of $\sim 3$ \citep{ggg+08}. 

The burst search algorithm that has successfully
discovered X-ray bursts from several AXPs and \kes\ did not reveal any 
bursts from the direction of \psr. We searched 751\,ks of \rxte\
observations for X-ray bursts, implying that the burst rate must be less 
than one burst per 751\,ks. This is less than the burst rate for \kes,
where 5 X-ray bursts have been observed, implying a burst rate of 
one per 407\,ks \citep{lnk+10}. 
{\it{Bona fide}} AXPs have a wide range of observed burst rates, from $<1$
burst per $\sim$700\,ks in 1RXS J170849.0$-$400910, 
up to 1 burst per $\sim$30\,ks for 1E~2259+586.

Finally, a study of the X-ray pulse profile
of \psr\ shows no significant variability on time scales of days to
years. Variations in individual pulse profiles are constrained to be 
less than 2.1\% from the average profile at the 3$\sigma$ level. 
Together, these results confirm the hypothesis that RPPs
can have stable X-ray properties on both 
short- and long-term time scales.

\subsection{Radiative variability from rotation-powered pulsars}
Previous studies have shown a correlation between $\dot{E}$ and $L_X$ in the soft
X-ray band for magnetospheric emission from RPPs and their nebular emission
\citep{sw88,bt97,kp08}. This can be interpreted as a constant value of the
conversion of spin-down luminosity into X-ray luminosity, 
$\eta_X=L_X/\dot{E}$, for all RPPs. Establishing 
a similar correlation between pulsed X-ray luminosity and 
$\dot{E}$ would be interesting, and implicitly requires $L_X$ to be stable. Prior to this
study of \psr, the long-term stability of $L_X$ had not been well established 
for any pulsar, largely owing to systematic uncertainties when comparing
results from different telescopes. 
Here, we show that $L_X$ can be steady on day-to-year-long time scales
for RPPs. This confirms the initial hypothesis, and stands in stark
contrast to the magnetars, where $L_X$ varies dramatically. 

Recent studies have shown that radio emission variability is
more common in pulsars than previously thought
\citep[e.g.][]{lhk+10,wje11}.
Specifically, the radio emission variability seen in the
``intermittent'' pulsars \citep{klo+06,lhk+10} appears to be
associated with large-scale magnetospheric variations.
If such a mechanism were
active in \psr, variations in $L_X$ would be expected, and should be
visible in our data, along with variations in the X-ray pulse profile.
Because
high-energy pulsar emission is more closely associated with the outer
magnetosphere, naively, one might  expect more dramatic variations in
the X-ray regime than in the energetically unimportant radio regime.
So far this cannot be confirmed, because 
none of the ``intermittent'' pulsars have detectable
magnetospheric X-ray pulsations, and no variations in the X-ray pulse
profile of \psr\ were detected here. 

Interestingly, two high $B$-field RPPs, \kes\ and J1119$-$6127, 
have shown unusual radiative behavior
near the time of a rotational glitch, in the X-ray and radio regimes,
respectively \citep{kh09,lkg10,wje11}. Given
this apparent propensity for pulsars with higher-than-average 
$B$-fields to display
radiative variability and \psr's relatively large magnetic field of
$1.5\times10^{13}$\,G, future 
observations of both the radio and X-ray emission properties of \psr\ may
show unusual behavior, particular in the case of a glitch. 

\subsection{Relationship between $B$ and magnetar outbursts}

The exact relationship between the estimated dipole field strength and
the ability and/or frequency of a neutron star to exhibit magnetar
activity remains to be established. However, there is now a wide range
of dipole $B$ estimates in neutron stars exhibiting magnetar
outbursts. 

\citet{pp11} suggest that all neutron stars may exhibit the type of
X-ray 
outbursts typically associated with magnetars, but that the frequency
of events is highly dependent on both the age and the magnetic field
strength of the source. For very young neutron stars ($<$2\,kyr),
outbursts may be observed roughly yearly for $B\simeq8\times10^{14}$\,G,
but only every 50--100\,yr for $B\simeq2 \times10^{14}$\,G. Thus for
neutron stars with $B\simeq 10^{13}$\,G, such as \psr\ and \kes,
outbursts should be less frequent than once per century. This
implies that the observation of an outburst from \kes\ was relatively
unlikely given that it has only been observed for $\sim$11\,yr, and
that 
the lack of magnetic activity from \psr\ is not surprising.
Ongoing observations of many high-$B$
radio pulsars should increase the chances of further observations of
magnetar-like bursts from RPPs. 

\subsection{Braking index, timing noise, and glitches}
\label{sec:timing_discussion}

We presented an extention of the timing solution of \psr\ from
21.3\,yr \citep{lkgm05} to 28.4\,yr. We found a braking index
of $n=2.832(3)$, 
which is 1.6$\sigma$ from the value based on 21.3\,yr of data.

\begin{figure}
\plotone{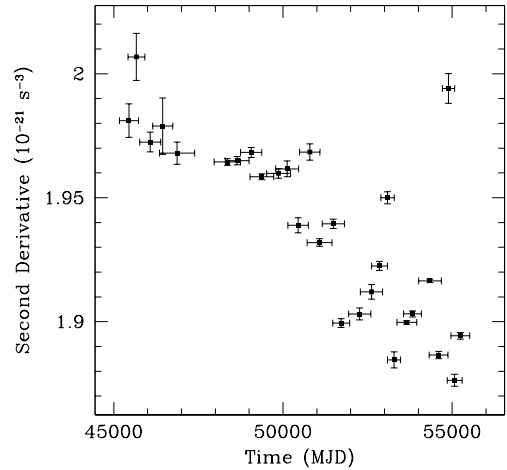}
\figcaption[]{Measurements of $\ddot\nu$ from short phase-coherent
timing solutions. Data prior to MJD~53000 were previously reported in
\citet{lkgm05}. A weighted least-squares fit to the data gives a slope
of $\nudotdotdot = 1.21(13) \times 10^{-31}$\,s$^{-4}$. The effect of
timing noise is visible as scatter in the data.
\label{fig:partial}}
\end{figure}

The scatter in short measurements of $\ddot\nu$, and therefore in $n$, is visible in 
both Figures~\ref{fig:partial} and \ref{fig:braking}, and is
typical of timing noise in young pulsars. We observed that the
root-mean-square variation in $n$ is $3.5\%$, larger than the
variations observed in the Crab pulsar of $\sim 0.5\%$ \citep{lps93}. 
We note that there is evidence for timing noise, but no evidence for a
glitch, in contrast to the suggestion made by \citet{hlk10} that timing
noise in pulsars with characteristic ages less than $\tau_c<10^{5}$\,yr
is entirely because of unmodeled glitch recovery. However, it remains
possible that \psr\ experiences very small glitches that cannot 
be distinguished from timing noise given the current data.

\normalsize

Because no glitches were found in the latest 7.1\,yr of timing
data, this extends the time without a detected glitch for \psr\ to 28.4\,yr.
While the youngest pulsars ($\tau_c<2$\,kyr) tend to have smaller
and fewer glitches than somewhat older pulsars ($\tau_c \sim
5-10$\,kyr), \psr\ is extreme in its lack of glitches. For example, 
the similarly aged Crab pulsar has had 24 glitches in 42 yr of
observations, PSR B0540$-$69 has had one glitch in 7.6\,yr, 
and PSR J1119$-$6127 has had 3 glitches in 12\,yr
\citep{elsk11,lkg05,wje11}. 

A higher order analog to the braking index
($n = {\ddot{\nu}}{\nu}/{\dot{\nu}^2}$), 
is the ``second braking index'', which can be predicted from a
measurement of $n$ to be $m_0 \equiv n(2n-1)$ and measured with a value 
of $\nudotdotdot$ to be
$m = {\nudotdotdot\nu^2}/{\dot\nu}^3$. Timing noise and glitches
prevent a measurement of $n$ for most pulsars. A measurement of $m$
is even more rare -- significant measurements have only been made for
the Crab pulsar and \psr\ \citep{lps93,kms+94,lkgm05}. If the standard
spin-down model ($\dot\nu=-K\nu^n$) is an accurate description of the
spin down of the pulsar, then $m=m_0$. 
By contrast, $n<3$ and $m\neq m_0$ implies that $K=K(t)$ and 
a measurement
of $m$ different from $m_0$ provides constraints on the functional form of
$K(t)$ \citep{br88}. 

Using the full span of available data of 28.4\,yr, we found
$\nudotdotdot = -1.21(13)\times10^{-31}$\,s$^{-4}$, which is 
within 1$\sigma$ from the previously reported value based on 21.3\,yr
of data. For the current measurements of $n$ and $\nudotdotdot$, we find 
that $m_0=13.21\pm0.03$ and $m=17.6\pm1.9$, which 
differ at the 2.3$\sigma$ level (see Table~\ref{table:coherent} for all 
rotational parameters). If uncertainties continue to decrease
at the same rate as over the last 7.1\,yr since the previous
measurement, and the measured value of $m$ remains stable, with
another $\sim$6\,yr of timing observations, the uncertainties should be
small enough that $m$ would differ from $m_0$ at the 3$\sigma$ level. 
If this proves to be the case and the two values are inconsistent,
this would
present challenges to the standard model of pulsar 
spin-down \citep[e.g., ][]{br88,bla94}. 

\section{Conclusions}

Analyzing 14.7\,yr of \rxte\ data, we found no evidence of variability 
in the X-ray emission of \psr. We have placed limits on possible
changes in the pulsed flux, profile variability, and burst rate from
\psr. This indicates that it is indeed possible, as typically assumed, 
for the X-ray emission properties of RPPs to be stable on day to decade long
time scales. 

Several RPPs have dipole $B$ estimates close
to, or larger than, magnetar-strength fields. These RPPs are typically
referred to as the high $B$-field pulsars. The link between
these and magnetars is not well understood and is under
investigation \citep[e.g.][]{oklk10,zkm+11,zkgl09,nk10}.
These pulsars, along with \kes, are possibly related to the
transient  magnetars \citep{km05}, such as XTE~1810$-$197
which experience long periods of quiescence,
followed  by magnetar activity \citep{ims+04}.
  
If the internal $B$ field of a neutron star can be much larger
than the observed external field responsible for the spin-down 
\citep[as suggested for SGR~0418+5729;][]{ret+10}
then it becomes difficult to predict which pulsars may experience
a future magnetar-like outburst. Indeed, long-term monitoring
observations of \kes\ showed that a sudden and
dramatic (though temporary)
change from a pulsar to a magnetar is possible.
Thus X-ray monitoring of pulsars such as that performed 
with \xte\ of \psr\ provides an opportunity to discover             
future outbursts from less-than-obvious magnetar candidates.

\acknowledgments
This research made use of data obtained from the High
Energy Astrophysics Science Archive Research Center Online Service, provided
by the NASA-Goddard Space Flight Center. VMK holds a Canada Research Chair and
the Lorne Trottier Chair in Cosmology and Astrophysics. 
Funding support for this work was provided by 
an NSERC Discovery Grant and Polanyi Prize, from FQRNT (Fonds
quebecois de la recherche sur la nature et les technologies) via the
Centre de Recherche en Astrophysique du Qu\'ebec (CRAQ), CIFAR (Canadian
Institute for Advanced Research) and from a Killam Research
Fellowship. 

\begin{deluxetable}{lc}
\tablecaption{X-ray Spin Parameters for \psr\ \label{table:coherent}}
\tablewidth{0pt}
\startdata
\cutinhead{Parameters for phase-coherent analysis}
Dates (Modified Julian Day)         & 50148.096~--~55521.082 \\
Epoch (Modified Julian Day)         & 52834.589000    \\
N$_{\rm TOA}$                       & 188             \\
$\nu$ (Hz)                            & 6.611515243850(3) \\
$\dot{\nu}$ ($10^{-11}$~s$^{-2}$)     & $-$6.694371307(6) \\
$\ddot{\nu}$ ($10^{-21}$~s$^{-3}$)    & 1.9185594(5)      \\
$\nudotdotdot$ ($10^{-31}$~s$^{-4}$)  & $-$0.9139(2)      \\ 
\cutinhead{Parameters for partially coherent analysis}
Average Braking Index, $n$           & 2.832(3)  \\
$m_0 \equiv n(2n-1)$                   & 13.21(3)   \\
$\nudotdotdot$ ($10^{-31}$~s$^{-4}$) & $-1.21(13)$ \\
Second braking index, $m$            & $17.6(1.9)$ \\
\enddata 
\tablecomments{Spin parameters for 14.7\,yr of X-ray timing data for \psr.}
\end{deluxetable}

\begin{deluxetable}{lcccc}
\tablecaption{Partially coherent ephemerides \label{table:partial}}
\tablewidth{0pt}
\tablehead{\colhead{Epoch}&\colhead{$\nu$}  &\colhead{$\dot\nu$} & \colhead{$\ddot\nu$} & \colhead{Date range} \\
\colhead{(MJD)}&\colhead{(Hz)}&\colhead{($10^{-11}$\,Hz$^{-2}$)}&\colhead{($10^{-21}$\,Hz$^{-3}$)}&\colhead{(MJD)}}
\startdata
50448 & 6.6253600487(2) &$-$6.734167(2)   &1.939(3)  &50148--50749\\
51083 & 6.6216683438(2) &$-$6.7234900(8)  &1.932(2)  &50750--51449\\
51853 & 6.6171996051(2) &$-$6.710680(1)   &1.917(3)  &51470--52167\\
52616 & 6.6127798918(2) &$-$6.698028(1)   &1.912(3)  &52265--52926\\
53091 & 6.6100326310(1) &$-$6.6901006(7)  &1.950(2)  &52893--53289\\
53647 & 6.60682103661(5)&$-$6.6809030(3)  &1.9032(6) &53311--53982\\
54345 & 6.60279542362(6)&$-$6.6694306(4)  &1.9165(7) &54007--54681\\
54892 & 6.5996455256(2) &$-$6.660372(2)   &1.994(6)  &54707--55076\\
55241 & 6.5976380501(1) &$-$6.6546543(5)  &1.894(1)  &54960--55522\\
\enddata
\tablecomments{Uncertainties are formal \tempo\ uncertainties. These
spin parameters are used for the pulsed flux analysis; see
Section~\ref{sec:flux}.} 
\end{deluxetable}

\begin{deluxetable}{lccc}
\tablecaption{Minimum burst fluences to detect bursts
\label{table:bursts}}
\tablewidth{0pt}
\tablehead{\colhead{Time resolution}&\colhead{Total
counts}& \colhead{Counts/PCU}&\colhead{Burst fluence} \\
\colhead{(ms)} & \colhead{} & \colhead{} &
\colhead{($10^{-12}$\,erg\,cm$^{-2}$)}}
\startdata
15.625 & 20 & 7  & 2.6 \\ 
31.25  & 25 & 8  & 5.9\\
62.25  & 30 & 10 & 15\\
125.0  & 40 & 13 & 39  \\
\enddata
\tablecomments{All fluences are for the 2--20\,keV energy range. 
Fluences are determined from the simulations described in 
Section~\ref{sec:bursts}. }
\end{deluxetable}


\begin{thebibliography}{44}
\expandafter\ifx\csname natexlab\endcsname\relax\def\natexlab#1{#1}\fi

\bibitem[{Becker \& Tr{\"u}mper(1997)}]{bt97}
Becker, W. \& Tr{\"u}mper, J. 1997, A\&A, 326, 682

\bibitem[{Blandford(1994)}]{bla94}
Blandford, R.~D. 1994, MNRAS, 267, L7

\bibitem[{Blandford \& Romani(1988)}]{br88}
Blandford, R.~D. \& Romani, R.~W. 1988, MNRAS, 234, 57P

\bibitem[{{Cusumano} {et~al.}(2001){Cusumano}, {Mineo}, {Massaro},
{Nicastro}, {Trussoni}, {Massaglia}, {Hermsen}, \& {Kuiper}}]{cmm+01}
{Cusumano}, G., {Mineo}, T., {Massaro}, E., {Nicastro}, L.,
{Trussoni}, E., {Massaglia}, S., {Hermsen}, W., \& {Kuiper}, L. 2001, A\&A, 375, 397
    
\bibitem[{{Efron}(1979)}]{efr79}
{Efron}, B. 1979, The Annals of Statistics, 7, 1
    
\bibitem[{{Espinoza} {et~al.}(2011){Espinoza}, {Lyne}, {Stappers},
 \& {Kramer}}]{elsk11}
{Espinoza}, C.~M., {Lyne}, A.~G., {Stappers}, B.~W., \&
{Kramer}, M. 2011, \mnras, 414, 1679
	
\bibitem[{{Gaensler} {et~al.}(1999){Gaensler}, {Brazier}, {Manchester},
  {Johnston}, \& {Green}}]{gbm99}
{Gaensler}, B.~M., {Brazier}, K.~T.~S., {Manchester}, R.~N.,
{Johnston}, S., \& {Green}, A.~J. 1999, \mnras, 305, 724
  
\bibitem[{{Gavriil} {et~al.}(2008){Gavriil}, {Gonzalez}, {Gotthelf},
{Kaspi}, {Livingstone}, \& {Woods}}]{ggg+08}
{Gavriil}, F.~P., {Gonzalez}, M.~E., {Gotthelf}, E.~V., {Kaspi},
 V.~M., {Livingstone}, M.~A., \& {Woods}, P.~M. 2008, Science, 319, 1802
      
\bibitem[{Gavriil {et~al.}(2004)Gavriil, Kaspi, \& Woods}]{gkw04}
Gavriil, F.~P., Kaspi, V.~M., \& Woods, P.~M. 2004, ApJ, 607, 959
      
\bibitem[{{Hobbs} {et~al.}(2010){Hobbs}, {Lyne}, \&
{Kramer}}]{hlk10}
{Hobbs}, G., {Lyne}, A.~G., \& {Kramer}, M. 2010, \mnras, 402, 1027
      
\bibitem[{{Ibrahim} {et~al.}(2004){Ibrahim}, {Markwardt},
 {Swank}, {Ransom},{Roberts}, {Kaspi}, {Woods}, {Safi-Harb}, {Balman}, {Parke},
{Kouveliotou}, {Hurley}, \& {Cline}}]{ims+04}
 {Ibrahim}, A.~I., {Markwardt}, C.~B., {Swank}, J.~H.,
 {Ransom}, S., {Roberts}, M., {Kaspi}, V., {Woods}, P.~M., {Safi-Harb}, S.,
 {Balman}, S., {Parke}, W.~C., {Kouveliotou}, C., {Hurley}, K., \& {Cline}, T.
  2004, ApJ, 609, L21
	      
\bibitem[{{Jahoda} {et~al.}(2006){Jahoda}, {Markwardt},
 {Radeva}, {Rots}, {Stark}, {Swank}, {Strohmayer}, \& {Zhang}}]{jmr+06}
{Jahoda}, K., {Markwardt}, C.~B., {Radeva}, Y.,
{Rots}, A.~H., {Stark}, M.~J., {Swank}, J.~H., {Strohmayer}, T.~E., \& {Zhang}, W.
 2006, ApJS, 163, 401
		  
\bibitem[{Jahoda {et~al.}(1996)Jahoda, Swank, Giles, Stark,
Strohmayer, Zhang, \& Morgan}]{jsg+96}
Jahoda, K., Swank, J.~H., Giles, A.~B., Stark, M.~J., Strohmayer,
 T., Zhang, W., \& Morgan, E.~H. 1996, Proc. SPIE, 2808, 59
    
\bibitem[{{Kargaltsev} \& {Pavlov}(2008)}]{kp08}
{Kargaltsev}, O. \& {Pavlov}, G.~G. 2008, in American Institute of
Physics Conference Series, Vol. 983, 40 Years of Pulsars: Millisecond
Pulsars, Magnetars and More, ed. {C.~Bassa, Z.~Wang, A.~Cumming, \&
V.~M.~Kaspi}, 171--185
	  
\bibitem[{Kaspi(2007)}]{kas07}
Kaspi, V.~M. 2007, Astrophys. Space Sci., 308, 1
	  
\bibitem[{{Kaspi}(2010)}]{kas10}
 {Kaspi}, V.~M. 2010, Proceedings of the National Academy of
  Science, 107, 7147
	  
\bibitem[{Kaspi {et~al.}(1994)Kaspi, Manchester, Siegman,
  Johnston, \& Lyne}]{kms+94}
 Kaspi, V.~M., Manchester, R.~N., Siegman, B., Johnston,
 S., \& Lyne, A.~G. 1994, ApJ, 422, L83
	      
\bibitem[{{Kaspi} \& {McLaughlin}(2005)}]{km05}
 {Kaspi}, V.~M. \& {McLaughlin}, M.~A. 2005, ApJ, 618

\bibitem[{{Kramer} {et~al.}(2006){Kramer}, {Lyne}, {O'Brien},
{Jordan}, \&  {Lorimer}}]{klo+06}
 {Kramer}, M., {Lyne}, A.~G., {O'Brien}, J.~T., {Jordan}, C.~A., \&
 {Lorimer}, D.~R. 2006, Science, 312, 549
    
\bibitem[{{Kuiper} \& {Hermsen}(2009)}]{kh09}
{Kuiper}, L. \& {Hermsen}, W. 2009, A \& A, 501, 1031
    
\bibitem[{{Kumar} \& {Safi-Harb}(2008)}]{ks08a}
{Kumar}, H.~S. \& {Safi-Harb}, S. 2008, ApJL, 678, L43
    
\bibitem[{{Livingstone} {et~al.}(2005{\natexlab{a}}){Livingstone},
 {Kaspi}, \& {Gavriil}}]{lkg05}
 {Livingstone}, M.~A., {Kaspi}, V.~M., \& {Gavriil}, F.~P.
   2005{\natexlab{a}}, ApJ, 633, 1095

\bibitem[{{Livingstone} {et~al.}(2010){Livingstone}, {Kaspi}, \&
  {Gavriil}}]{lkg10}
  ---. 2010, ApJ, 710, 1710

\bibitem[{{Livingstone}	{et~al.}(2005{\natexlab{b}}){Livingstone}, {Kaspi},
  {Gavriil}, \& {Manchester}}]{lkgm05}
  {Livingstone}, M.~A., {Kaspi}, V.~M., {Gavriil}, F.~P., \&
  {Manchester}, R.~N. 2005{\natexlab{b}}, ApJ, 619, 1046
	    
\bibitem[{{Livingstone} {et~al.}(2011){Livingstone}, {Ng},
    {Kaspi}, {Gavriil}, \& {Gotthelf}}]{lnk+10}
{Livingstone}, M.~A., {Ng}, C.-Y., {Kaspi}, V.~M.,
 {Gavriil}, F.~P., \& {Gotthelf}, E.~V. 2011, \apj, 730, 66
		
\bibitem[{{Lyne} {et~al.}(2010){Lyne}, {Hobbs},
{Kramer}, {Stairs}, \&  {Stappers}}]{lhk+10}
 {Lyne}, A., {Hobbs}, G., {Kramer}, M., {Stairs}, I.,
 \& {Stappers}, B. 2010, Science, 329, 408

\bibitem[{Lyne {et~al.}(1993)Lyne, Pritchard, \& Smith}]{lps93}
Lyne, A.~G., Pritchard, R.~S., \& Smith, F.~G. 1993, MNRAS, 265, 1003

\bibitem[{Manchester \& Taylor(1977)}]{mt77}
Manchester, R.~N. \& Taylor, J.~H. 1977, Pulsars (San Francisco:
Freeman)

\bibitem[{Marsden {et~al.}(1997)Marsden, Blanco, Gruber, Heindl,
Pelling, Peterson, Rothschild, Rots, Jahoda, \& Macomb}]{mbg+97}
Marsden, D., Blanco, P.~R., Gruber, D.~E., Heindl, W.~A., Pelling,
M.~R., Peterson, L.~E., Rothschild, R.~E., Rots, A.~H., Jahoda, K., \&
Macomb, D.~J. 1997, ApJ, 491, L39
      
\bibitem[{{Mereghetti}(2008)}]{mer08}
 {Mereghetti}, S. 2008, \aapr, 15, 225
      
\bibitem[{{Ng} \& {Kaspi}(2010)}]{nk10}
  {Ng}, C.~Y. \& {Kaspi}, V.~M. 2010, arXiv:1010.4592
     
\bibitem[{{Olausen} {et~al.}(2010){Olausen}, {Kaspi}, {Lyne}, \&
   {Kramer}}]{oklk10}
{Olausen}, S.~A., {Kaspi}, V.~M., {Lyne}, A.~G., \& {Kramer},
M. 2010, \apj, 725, 985
	  
\bibitem[{{Perna} \& {Pons}(2011)}]{pp11}
  {Perna}, R. \& {Pons}, J.~A. 2011, \apjl, 727, L51+

\bibitem[{{Rea} \& {Esposito}(2011)}]{re11}
{Rea}, N. \& {Esposito}, P. 2011, in High-Energy Emission from Pulsars
and their Systems, eds. {D.~F.~Torres \& N.~Rea}, 247
  
\bibitem[{{Rea} {et~al.}(2010){Rea}, {Esposito}, {Turolla},
 {Israel}, {Zane}, {Stella}, {Mereghetti}, {Tiengo}, {G{\"o}tz}, 
 {G{\"o}{\u g}{\"u}{\c s}}, \& {Kouveliotou}}]{ret+10}
 {Rea}, N., {Esposito}, P., {Turolla}, R., {Israel}, G.~L.,
 {Zane}, S., {Stella}, L., {Mereghetti}, S., {Tiengo}, A., {G{\"o}tz}, D.,
{G{\"o}{\u g}{\"u}{\c s}}, E., \& {Kouveliotou}, C. 2010, Science, 330, 944
	  
\bibitem[{Rots {et~al.}(1998)Rots, Jahoda, Macomb, Kawai,
  Saito, Kaspi, Lyne, Manchester, Backer, Somer, Marsden, \& Rothschild}]{rjm+98}
  Rots, A.~H., Jahoda, K., Macomb, D.~J., Kawai, N., Saito,
  Y., Kaspi, V.~M., Lyne, A.~G., Manchester, R.~N., Backer, D.~C., Somer,
   A.~L., Marsden, D., \&  Rothschild, R.~E. 1998, ApJ, 501, 749
		
\bibitem[{Seward \& Harnden~Jr.(1982)}]{sh82}
Seward, F.~D. \& Harnden~Jr., F.~R. 1982, ApJ, 256, L45
		
\bibitem[{Seward \& Wang(1988)}]{sw88}
Seward, F.~D. \& Wang, Z.-R. 1988, ApJ, 332, 199
		
\bibitem[{Thompson \& Duncan(1996)}]{td96a}
Thompson, C. \& Duncan, R.~C. 1996, ApJ, 473, 322

\bibitem[{Thompson {et~al.}(2002)Thompson, Lyutikov, \&
Kulkarni}]{tlk02}
Thompson, C., Lyutikov, M., \& Kulkarni, S.~R. 2002, ApJ, 574, 332

\bibitem[{{Weltevrede} {et~al.}(2011){Weltevrede}, {Johnston}, \&
  {Espinoza}}]{wje11}
  {Weltevrede}, P., {Johnston}, S., \& {Espinoza}, C.~M. 2011, \mnras,
  411, 1917
  
\bibitem[{Woods \& Thompson(2006)}]{wt06}
Woods, P.~M. \& Thompson, C. 2006, in Compact Stellar X-ray Sources,
 ed. W.~H.~G. Lewin \& M.~van~der Klis (UK: Cambridge University Press)
    
\bibitem[{{Zhu} {et~al.}(2009){Zhu}, {Kaspi}, {Gonzalez}, \&
{Lyne}}]{zkgl09}
 {Zhu}, W., {Kaspi}, V.~M., {Gonzalez}, M.~E., \& {Lyne}, A.~G.
 2009, ApJ, 704, 1321
      
 \bibitem[{{Zhu} {et~al.}(2011){Zhu}, {Kaspi}, {McLaughlin},
 {Pavlov}, {Ng}, {Manchester}, {Gaensler}, \& {Woods}}]{zkm+11}
{Zhu}, W.~W., {Kaspi}, V.~M., {McLaughlin}, M.~A., {Pavlov},
G.~G., {Ng}, C.-Y., {Manchester}, R.~N., {Gaensler}, B.~M., \& {Woods},
  P.~M. 2011, \apj, 734, 44
	    
\end{thebibliography}
\end{document}